\documentclass[conference]{IEEEtran}
\IEEEoverridecommandlockouts
\usepackage{cite}
\usepackage{amsmath,amssymb,amsfonts}
\usepackage{algorithmic}
\usepackage{graphicx}
\usepackage{textcomp}
\usepackage{xcolor}
\usepackage{booktabs}
\usepackage{subcaption}
\usepackage{url}
\def\BibTeX{{\rm B\kern-.05em{\sc i\kern-.025em b}\kern-.08em
    T\kern-.1667em\lower.7ex\hbox{E}\kern-.125emX}}
\begin{document}

\title{Understanding Large-Scale HPC System Behavior \\Through Cluster-Based Visual Analytics
% {\footnotesize \textsuperscript{*}Note: Sub-titles are not captured for https://ieeexplore.ieee.org  and
}

% \author{
%     \IEEEauthorblockN{Allison Austin}
%     \IEEEauthorblockA{\textit{University of California, Davis}\\ Davis, CA, USA \\ amaustin@ucdavis.edu}
%     \and
%     \IEEEauthorblockN{Shilpika}
%     \IEEEauthorblockA{\textit{Argonne National Laboratory}\\ Lemont, IL, USA\\shilpika@anl.gov}
%     \and
%     \IEEEauthorblockN{Yan To Linus Lam}
%     \IEEEauthorblockA{\textit{University of California, Davis}\\ Davis, CA, USA\\ytllam@ucdavis.edu}
%     \and 
%     \IEEEauthorblockN{Yun-Hsin Kuo}
%     \IEEEauthorblockA{\textit{University of California, Davis}\\ Davis, CA, USA\\yskuo@ucdavis.edu}
%     \and
%     \IEEEauthorblockN{Venkatram Vishwanath}
%     \IEEEauthorblockA{\textit{Argonne National Laboratory}\\ Lemont, IL, USA\\venkat@anl.gov}
%     \and 
%     \IEEEauthorblockN{Michael E. Papka}
%     \IEEEauthorblockA{\textit{Argonne National Laboratory}\\ \textit{University of Illinois Chicago}\\ Illinois, USA\\papka@anl.gov}
%     \and 
%     \IEEEauthorblockN{Kwan-Liu Ma}
%     \IEEEauthorblockA{\textit{University of California, Davis}\\ Davis, CA, USA\\klma@ucdavis.edu}
% }

\author{
    \IEEEauthorblockN{
        Allison Austin\IEEEauthorrefmark{1},
        Shilpika\IEEEauthorrefmark{2},
        Yan To Linus Lam\IEEEauthorrefmark{1},
        Yun-Hsin Kuo\IEEEauthorrefmark{1},\\
        Venkatram Vishwanath\IEEEauthorrefmark{2},
        Michael E. Papka\IEEEauthorrefmark{2}\IEEEauthorrefmark{3}, 
        and Kwan-Liu Ma\IEEEauthorrefmark{1}
    }
    \IEEEauthorblockA{
        \IEEEauthorrefmark{1}University of California, Davis, Davis, CA, USA \\ \{amaustin, ytllam, yskuo, klma\}@ucdavis.edu\\
        \IEEEauthorrefmark{2}Argonne National Laboratory, Lemont, IL, USA \\ \{shilpika, venkat, papka\}@anl.gov\\
        \IEEEauthorrefmark{3}University of Illinois Chicago, Chicago, IL, USA
    }
}

\maketitle

\begin{abstract}
In high-performance computing (HPC) environments, system monitoring data is often unlabeled and high-dimensional, making it difficult to reliably detect and understand anomalous computing nodes. The growing scale and dimensionality of the collected datasets present significant challenges for analysis and visualization tasks. We present a scalable, interactive visual analytics system to support exploration, explanation, and comparison of compute node behaviors in HPC systems. Our approach integrates an analysis workflow combining two-phase dimensionality reduction with contrastive learning and multi-resolution dynamic mode decomposition to capture inter- and intra-cluster variations. These analyses are embedded in an interactive interface that enables users to explore clusters, compare temporal patterns, and iteratively refine hypotheses through customizable visual encodings and baselines. By integrating metrics such as CPU utilization and memory activity, the system offers a holistic view of large-scale system behavior. We demonstrate the utility of our tool through two case studies. In both cases, our system automatically identified meaningful node clusters and revealed subtle behavioral differences within and across node groups. Expert feedback confirmed the effectiveness of our tool in enhancing anomalous behavior detection and interpretation. Our work advances scalable visual analysis for HPC monitoring and has broader implications for cloud, edge computing, and distributed infrastructures where interpretability and behavior analysis are critical to operational efficiency.
\end{abstract}

\begin{IEEEkeywords}
%visualization, 
visual analytics, time-series, multidimensional, multivariate, clustering, HPC, performance, monitoring
\end{IEEEkeywords}

\section{Introduction}

High-performance computing (HPC) systems are essential for tackling complex problems in computational and physical sciences, as they enable researchers to perform large-scale simulations with remarkable speed and precision. These HPC systems offer massive parallelism through high memory bandwidth, low‐latency interconnects, thousands of connected compute nodes, and scalable storage, all of which contribute to their complexity. Effective multifidelity monitoring of these complex systems plays a key role in ensuring successful application runs while conserving computation time and resources. These systems typically employ integrated monitoring tools that collect metrics from physical sensors and buffer data at a high rate~\cite{ganglia,zabbix,nagios}. However, existing monitoring tools mainly visualize only basic system status. This limited information hinders the ability of experts to identify causal relationships behind issues such as data corruption, resource contention, or disk cache eviction \cite{performancevar}.

Existing anomaly detection methods often require parameter tuning and expert-defined baselines, which are difficult to derive from raw, unlabeled monitoring data. Several existing works apply visual analytics to synthetic multivariate HPC monitoring data \cite{dbscan-env} or attempt to ``inject" anomalies into real-world monitoring data \cite{wintermute}. However, the effectiveness of these methods is not always translated when applied to real-world data. Analyzing real system monitoring data presents several unique challenges, including (but not limited to):

\begin{itemize}
    \item \textbf{Hidden hardware failures.} Hardware failure in HPC systems can often go undetected, leading to latent %cascading %root 
    issues that cause back pressure on the system and further performance degradation over time \cite{failslow}. Thus, anomalies need to be monitored actively and detected quickly.
    \item \textbf{Complex dependencies.} Causes of failures in HPC systems can be difficult to trace, as failures in one part of the system (CPU, network, I/O, scheduling) can cause cascading faults. 
    \item \textbf{Data variability and scale.} HPC monitoring data is often high-dimensional, noisy, and highly variable due to dynamic job scheduling, fluctuating resource usage, and environmental or operational noise. This variability makes it difficult for classical methods to distinguish anomalies from benign fluctuations, especially at scale.
    \item \textbf{No labeled data to support analysis.} Labeled ground truth—particularly for anomalies—is rarely available in production HPC systems. As a result, analysts rely on limited or proxy data to guide anomaly detection \cite{borghesi_anomaly_2022}. Contextual information from % job data and other 
    system logs is needed to isolate inherent system performance from external factors, such as maintenance scheduling or manual user intervention. However, this data is not always accessible or even collected in tandem with monitoring data. 
\end{itemize} 

To address these limitations, we design an interactive system % combining 
that combines visualization and analysis techniques to help users identify anomalies and behavior groups in dynamic, real-world HPC datasets. We %combine 
utilize both multivariate and univariate time series analysis techniques to generate explainable node clusters from HPC monitoring data and investigate intra- and inter-cluster differences. While each employed method individually supports only a subset of these analysis tasks, their combination enables forms of cluster-level and node-level temporal analysis that are not possible with any single technique alone. Our visualization serves as an abstraction layer, enabling users to interact with each method by adjusting input parameters such as time range, nodes, metrics, and baselines. These interactions are low-latency due to the system's built-in caching mechanism, allowing users to iteratively explore clustering, explanation, and temporal decomposition results without recomputing embeddings. This interaction allows expert analysts to refine the analysis based on prior knowledge to generate more precise and holistic results. 
Our visualization provides both a high-level overview of monitoring data and detailed node-level views, enabling analysts to quickly identify major system events without relying on raw system logs for context. Our system accelerates the process of investigating individual nodes with abnormal behavior and supports comparative analysis across groups of compute nodes and metrics (e.g., temperature, power, current) over different time periods.
Based on close collaboration with HPC experts and a literature review of monitoring visualizations, we derived key design requirements that guided our system development. The resulting tool runs remotely as a separate module, aggregating and processing data into a single, centralized view for interactive exploration. \\
The primary contributions of our work are as follows:
\begin{enumerate}
\item We %couple 
combine three state-of-the-art analysis techniques--two-phase dimensionality reduction (MulTiDR)~\cite{multidr}, contrastive learning (ccPCA)~\cite{ccpca}, and multi-resolution dynamic mode decomposition (mrDMD)~\cite{multifidelity}--into a unified workflow that enables both \textit{inter}-cluster and \textit{intra}-cluster analysis of multivariate time series data.
\item We design an intuitive, low-latency visual interface\footnote{\url{https://github.com/VIDILabs/node-cluster-vis}} that leverages caching of analysis % pipeline 
stages to support interactive, iterative exploration. This interface abstracts %pipeline 
workflow complexity while enabling rapid algorithm fine-tuning and configuration. 
\item We evaluate the system on real-world HPC monitoring datasets from two facilities, validating detected patterns against job logs and operational records. 
\end{enumerate}

This work demonstrates how advanced dimensionality reduction, contrastive analysis, and decomposition techniques can be combined with interactive visualization to improve HPC system anomaly detection and understanding at scale.
\section{Related Work}

Existing approaches for the analysis and visualization of HPC monitoring data focus on anomaly detection, performance tuning, and pattern recognition. 
Since we are concerned with explaining the behavior of unlabeled monitoring data, traditional machine learning techniques that require extensive pre-training \cite{tuncer_online_2019} and user-defined annotations are not well-suited to our tasks. 
Instead, dimensionality reduction and decomposition techniques % are often 
can be employed to handle multivariate, high-dimensional time series data collected from system components such as CPU, memory, and network metrics. Throughout this paper, we refer to the physical hardware as ``system" or ``compute/buffer nodes", and the term ``cluster" is reserved exclusively for \textit{behavioral groups} in the data space. 

\subsection{Dimensionality Reduction (DR)}
Methods such as Principal Component Analysis (PCA)~\cite{pca} and Uniform Manifold Approximation and Projection (UMAP)~\cite{umap} have shown significant promise in simplifying and interpreting high-dimensional, multivariate HPC monitoring datasets. These methods not only reduce computational complexity but also preserve critical structural information, enabling effective visualization and intuitive exploration of complex patterns and anomalies. Recent advancements in unsupervised data analysis techniques, particularly in Multi-resolution Dynamic Mode Decomposition (robust PCA at multiple resolutions)~\cite{mrdmd,multifidelity}, provide opportunities to model and interpret unseen monitoring behaviors without relying heavily on labeled datasets or extensive manual feature engineering. However, on their own, these techniques can sometimes be unreliable due to their sensitivity to parameter selection and their potential loss of important information during the compression process. Additionally, the interpretability of reduced dimensions can become challenging, as the transformed dimensions may not directly correspond to meaningful physical or system properties. Consequently, our work explores hybrid approaches that combine state-of-the-art advancements in DR with visual analytics techniques to achieve robust, interpretable, and scalable solutions for retrospective analysis of large-scale HPC monitoring data.

\subsection{Anomaly Detection in HPC Applications}
Numerous studies have explored anomaly detection in HPC systems, with early efforts focusing primarily on text-based logs~\cite{b33,surveyreliability,9197818}. 
These log analysis tools~\cite{b2,9825952, mela} use text pattern correlations, pattern recognition, and spatiotemporal text-based analysis. 
However, without log messages, analysis of system behavior must instead rely on large-scale time series data collected at different granularity levels--the lowest level typically being compute node sensor readings collected from integrated monitoring tools such as Ganglia \cite{ganglia}, Zabbix \cite{zabbix}, and Nagios \cite{nagios}, which buffer data from physical sensors at a high rate. 
Approaches for analyzing such time series data often target a single data type (e.g., jobs \cite{map,multifidelity,prodigy}, or network-level telemetry \cite{network-cngstn,parallel-apps,high-radix}), while low-level sensor data is frequently neglected due to its complexity. Analysis and visualization of sensor data is often paired with system logs or job data to provide context or ground truth \cite{mtsad,wintermute,prodigy,aksar_runtime_2024,borghesi_anomaly_2022,deephydra}. 
Recent work has begun to focus directly on analyzing this low-level sensor data~\cite{multifidelity,tak2019,incmrdmd,inc-msplot}, aiming to reveal system patterns and anomalies beyond what is captured in system logs. 
While these ML-based~\cite{tuncer_online_2019,ozer_characterizing_2020, ruad,das_prolego_2023,aksar_runtime_2024} and cluster-based~\cite{dbscan-env,multidr} approaches have shown promise in detecting anomalies in unlabeled sensor data, they face key limitations: lack of interpretability, high training cost, and limited configurability for end users. An empirical study of log analysis at Microsoft~\cite{10.1145/3540250.3558963} highlights this gap:\textit{“Anomaly detection results from a sophisticated model are often difficult to interpret, especially for engineers who do not understand machine learning.”} 
Our work precisely addresses this gap by abstracting out the more complicated details of the methodologies used, and our visualization pipeline drives the interpretability of the results. 

\subsection{Visual Analytics for HPC Monitoring}
Despite the progress demonstrated by existing visual analytics (VA) systems, none of the existing approaches we surveyed adequately support interpretable analysis of low-level multivariate sensor data without heavy assumptions or extensive preprocessing. Dashboard tools such as Grafana~\cite{grafana} provide only descriptive monitoring views and lack analytical capabilities such as behavioral clustering or comparative node analysis. Similarly, integrated monitoring tools \cite{ganglia,zabbix,nagios} offer basic summaries of system health and activity, but also lack advanced analytics, rich interactivity, and system-oriented insights. Recent VA systems and frameworks extend these capabilities by coupling scalable visualization with machine learning techniques \cite{mtsad,wintermute,prodigy}. However, these techniques rely on labeled data or offline training--limiting their applicability to environments where baseline node behavior is unknown or evolving.
Other operational data analytics (ODA) frameworks such as DCDB Wintermute~\cite{wintermute} and ExaMon~\cite{borghesi_anomaly_2022} provide unsupervised and semi-supervised detection, but their results do not expose intermediate reasoning that would enable human validation. 
Existing VA systems designed for interpreting and exploring multivariate time series data~\cite{mtsad,multidr,kesavan_visual_2020} are typically constrained to a small subset of variables (often fewer than 10 simultaneously) to avoid visual clutter and scalability issues. Systems that support high-dimensional multivariate time series data typically compress first and visualize after--i.e., they do not support interactive exploration of hundreds of raw variables in a centralized view. This is also largely due to the curse of dimensionality, wherein the contrast between distances of different data points diminishes as the number of dimensions considered in analysis increases.
As a result, these systems do not scale to high-dimensional system telemetry, where hundreds of metrics per node are common. 
This is the motivation for our exploratory approach on inter- and intra-cluster analysis with baseline-aware anomaly detection. 

Our work addresses the mentioned limitations by combining unsupervised behavior extraction and interactive visual analysis to produce explainable and user-configurable monitoring insights without requiring labeled data, retraining, or external log context. 
\section{Methodology}
\label{sec:methodology}
\subsection{Design Requirements}
To guide the design of our system, we collaborated with HPC experts from Argonne National Laboratory and Fermi National Accelerator Laboratory to identify five key analysis tasks that our system should support (\textbf{R1}--\textbf{R5}). These requirements informed both system design and feature prioritization, enabling analysts to efficiently explore and interpret large-scale HPC monitoring data. 

\begin{itemize}
\item[\bf R1.]{
{\bf Summarizing compute node behavior.} To support inter-cluster analysis at scale, the system should help users grasp the overall performance of hundreds of compute nodes.
}

\item[\bf R2.]{{\bf Depicting relationships among monitoring metrics.} Revealing the relationships among metrics is key to interpreting the cascading effects within the system, an essential aspect of diagnosing HPC performance. }

\item[\bf R3.]{{\bf 
Identifying anomalous nodes with explanations.} Irregular node behavior is often the root cause of anomalous events. Our system should highlight the compute nodes with significant behavioral deviations in certain monitoring metrics.
}

\item[\bf R4.]{{\bf
Providing system overview and raw data inspection.} Presenting overall runtime behavior and raw readings is essential for validating findings in large-scale HPC monitoring data.
}

\item[\bf R5.]{{\bf
Supporting transitions between inter-cluster and intra-cluster analysis.} Our system should support user interactions to help connect insights from both analysis components and build a coherent understanding of HPC system behavior.
}
\end{itemize}

\begin{figure}[ht]
    \centering
    \includegraphics[width=\columnwidth]{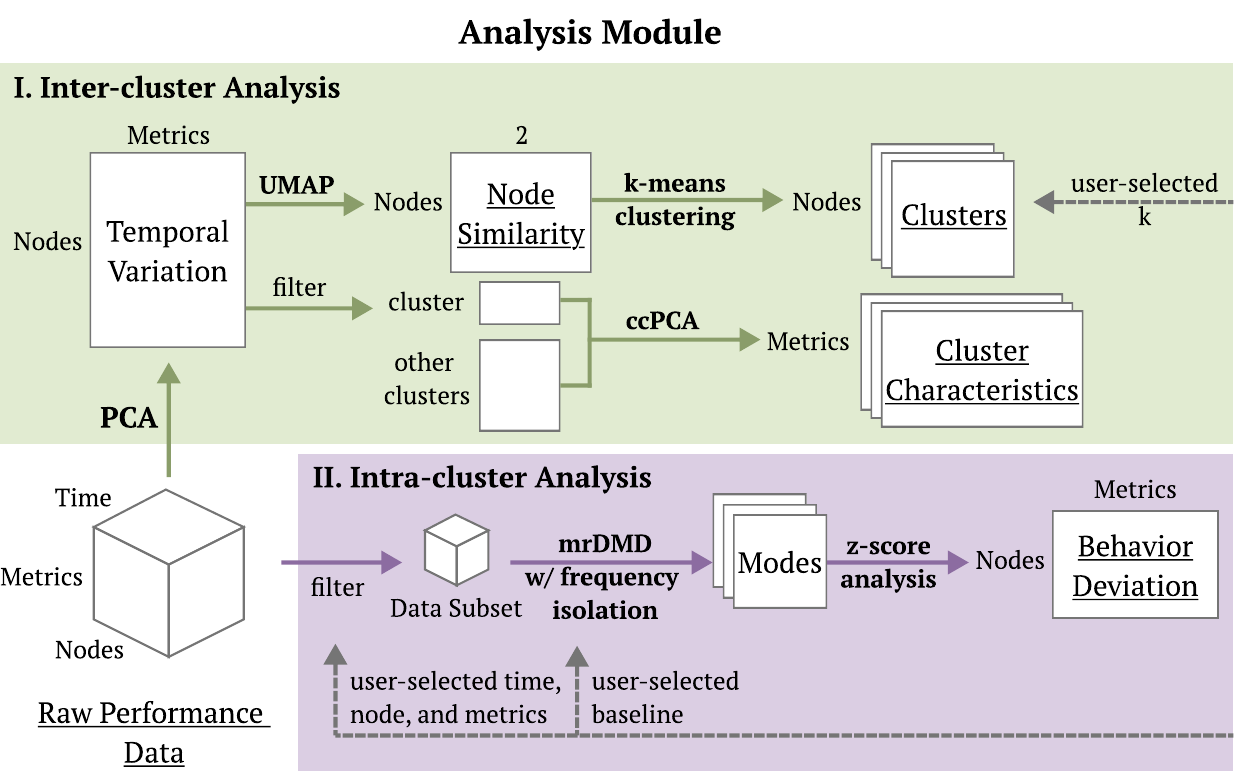}
    \caption{Analysis workflow of our system. Our analysis is driven by inter-cluster and intra-cluster analysis, where users interact with the system to refine the analysis results. Underlined data elements are visualized in the interface.}
    \label{fig:workflow}
    \vspace{-0.15in}
\end{figure}

\begin{figure*}[ht]
    \centering
    \includegraphics[width=0.9\linewidth]{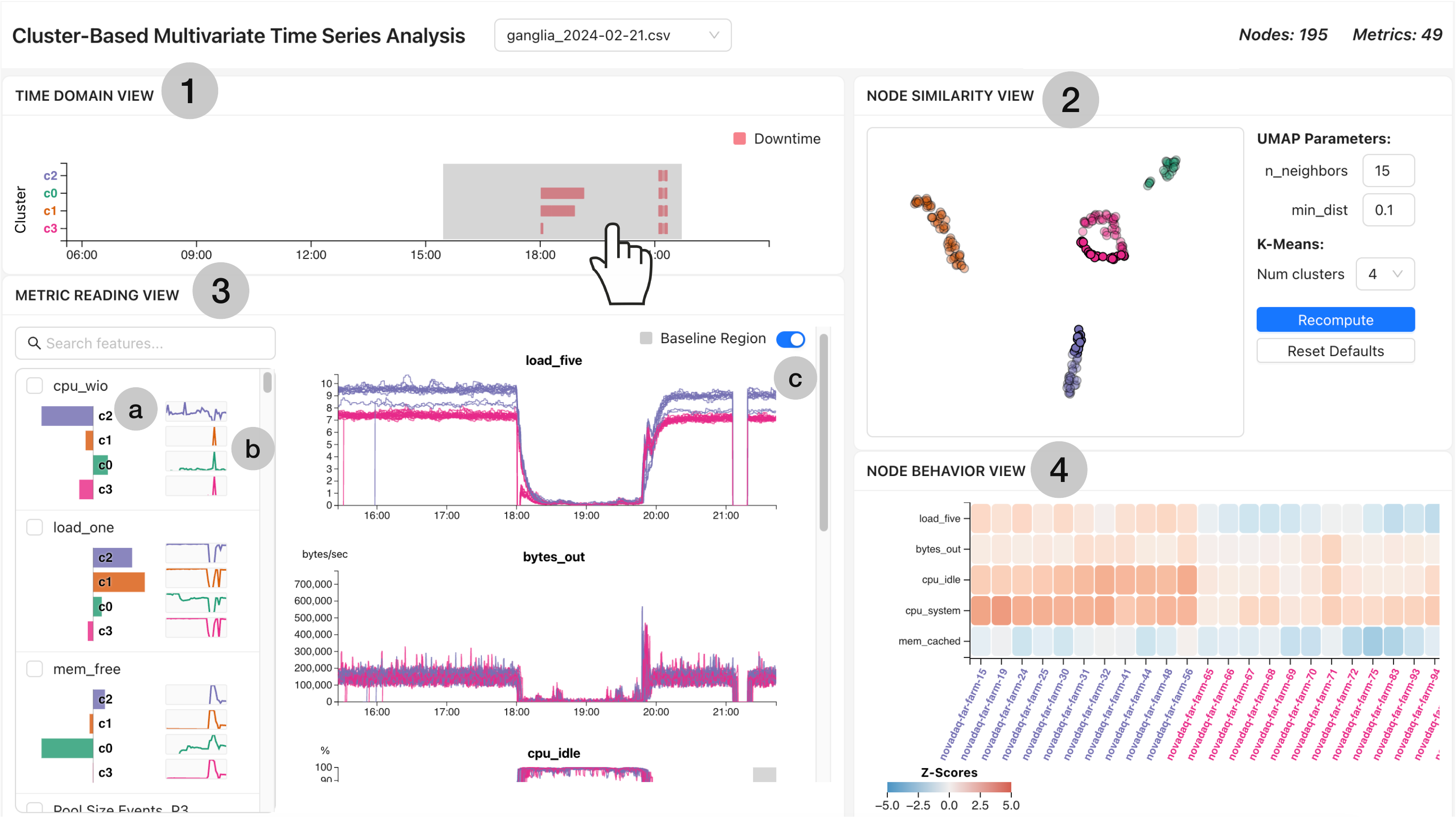}
    \caption{The system interface for intra- and inter-cluster analysis of multivariate HPC monitoring data. The visualization comprises four coordinated views: (1) Time Domain view for temporal navigation, (2) Node Similarity view displaying node clusters derived from UMAP applied to PCA results, (3) Metric Reading view showing cluster behavior across all metrics and raw time series for selected nodes, and (4) Node Behavior view presenting mrDMD z-scores. Brushing in the Time Domain view (1) controls the temporal range displayed in the time series plots (3c). Vertical bar charts in the metric list (3a) encode each metric's contribution to cluster formation. Polylines in (3b) represent cluster-averaged metric values over time.}
    \label{fig:vis}
    \vspace{-0.1in}
\end{figure*}

Our system operates on raw monitoring data capturing dynamic, multivariate readings at the node level, which can be conceptualized as a three-dimensional tensor along the node, time, and metric dimensions.
It comprises two modules: an analysis module (Fig. \ref{fig:workflow}) and a visualization module (Fig. \ref{fig:vis}).
In the following, we describe the analysis workflow and illustrate how the analysis is integrated into the visualization components.

\subsection{Analysis Module}

Identifying anomalous behavior is akin to slicing the data tensor along its dimensions into a smaller data cube as the analysis targets.
To support the exploration and analysis of large-scale performance data at interactive speed, we leverage cluster-based analysis, similar to chopping, to help system users drill down to their analysis targets of interest.
Our methodology is driven by two components: \textit{inter}-cluster analysis and \textit{intra}-cluster analysis, as depicted in Fig. \ref{fig:workflow}.
Inter-cluster analysis groups compute nodes based on distinct behavior patterns over time, offering system users an interpretable, high-level overview of raw performance data. 
Intra-cluster analysis incorporates user inputs to compare spatial and temporal variations within a cluster in detail, revealing anomaly nodes and facilitating interpretation of their behavior based on raw monitoring readings.
Together, these two levels of analysis allow users to alternate between overview and detail, supporting broad pattern discovery and fine-grained anomaly interpretation.

\subsubsection{Inter-cluster analysis}
The goal of inter-cluster analysis is to partition large-scale raw monitoring data into interpretable node clusters, thereby facilitating scalable data exploration.
Dimensionality reduction (DR) has been widely adopted in visual analytics to support cluster-based analysis, due to their effectiveness in capturing complex data structure~\cite{wenskovitch2017towards}.
By projecting high-dimensional data similarities into low-dimensional embeddings, typically visualized as 2D scatter plots, DR provides a succinct overview of complex data.
However, many DR methods take two-dimensional matrices as input, limiting the ability to capture interactions among the three dimensions of our tensor data.
We leverage MulTiDR~\cite{multidr}, a two-phase DR framework designed for analyzing tensor data. This approach stacks two tensor dimensions and compresses the remaining one in the first pass, producing a matrix suitable for the second pass.
Since we aim to uncover relationships among compute nodes over time, we first compress the time dimension, followed by the metric dimension.
Unlike stacking the metric and node dimensions to capture global patterns, our strategy is to iteratively compress each metric slice across time to preserve fine-grained temporal patterns.
We follow the DR configuration, PCA~\cite{pca} and UMAP~\cite{umap}, recommended by MulTiDR \cite{multidr}, for the former's capability in effective data compression and that of the latter in balanced neighborhood preservation.
Specifically, we use PCA to summarize each node's temporal pattern into a representative value of temporal variation.
Then, we apply UMAP to compress the temporal variation across metrics into a two-dimensional embedding, revealing the relationships among compute nodes that take both time and metric dimensions into account.
Finally, we cluster on the embedding to obtain the node groups. The clustering serves as a guide in our visualization system for user interactions based on the grouped nodes. We choose $k$-means clustering for its simplicity of parameter selection~\cite{algorithms}. We use the default parameters for UMAP (\textit{n\_neighbors}=15, \textit{min\_dist}=0.1), and a default $k$ of 4, but the clustering parameters and the baselines can all be tuned in the UI. 
The relationships among compute nodes, along with cluster labels, are presented in our Node Similarity view (Fig. \ref{fig:vis}-2) as a data overview (\textbf{R1}) and serve as the foundation for subsequent cluster-based analysis.

\begin{figure}[ht]
    \centering
    \includegraphics[width=0.9\columnwidth]{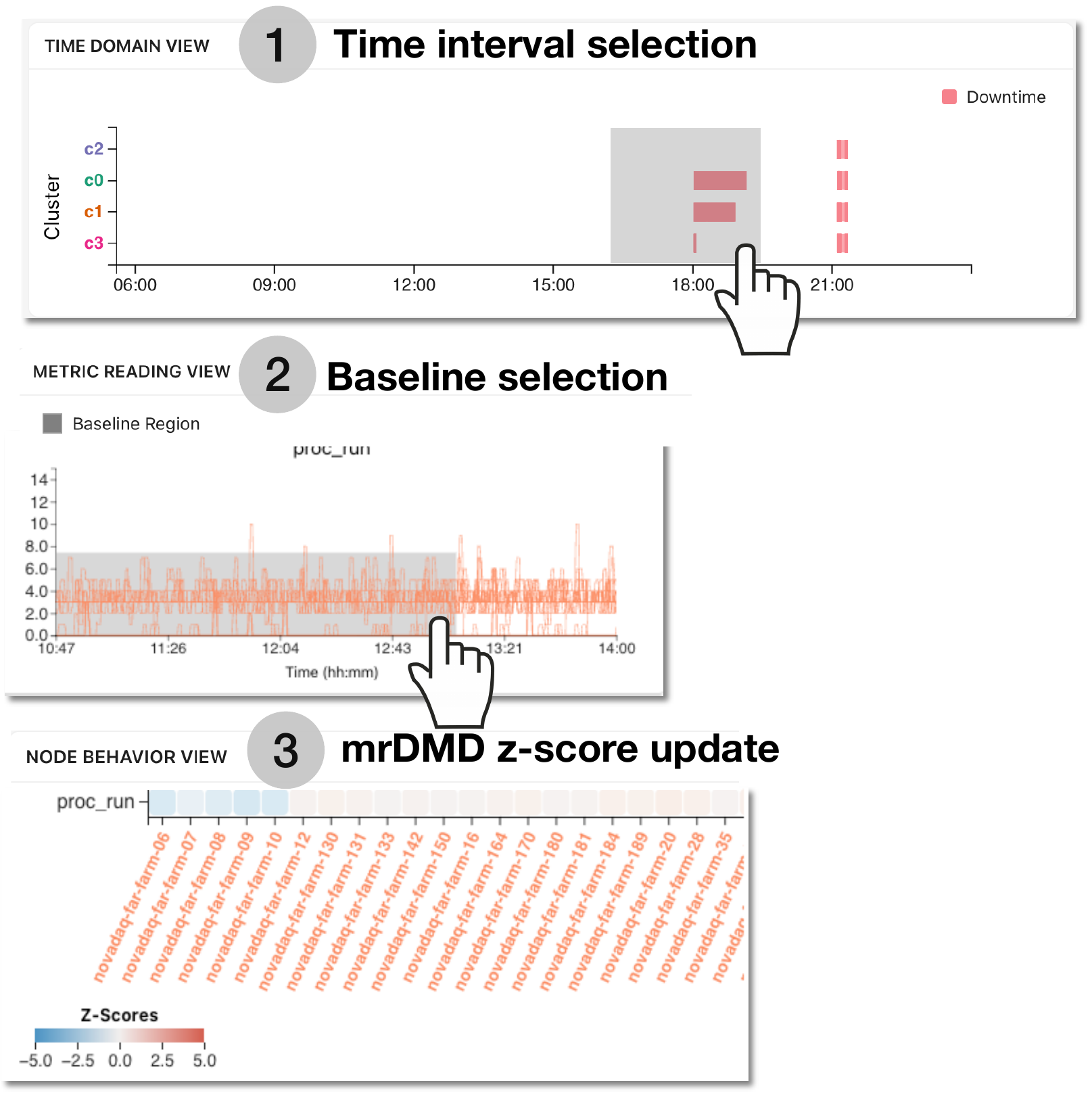}
    \caption{Visualization interaction with mrDMD z-score analysis. After adjusting the time window of interest (1), users can update the statistically-derived default baseline region (2) to recompute the mrDMD z-scores (3).}
    \label{fig:interact}
    \vspace{-0.1in}
\end{figure} 

To understand each node cluster and explain their differences, we characterize the node clusters based on their monitoring metrics. 
We employ a contrastive learning method, ccPCA~\cite{ccpca}, to support the interpretability of our Node Similarity view.
ccPCA takes two groups as input (a target group and a background group) and identifies metrics that are highly discriminative for the target group with respect to the background.
In our case, we partition the first-pass DR result (i.e., the input to the second-pass DR) based on the cluster labels to construct the input groups. 
To reveal each cluster's unique characteristics, we adopt the one-versus-rest approach; each cluster is treated in turn as the target group, with all other nodes forming the background.
This process yields a weight vector for each cluster, where each value reflects the contribution of each metric to the uniqueness of the target cluster (\textbf{R2}), as shown in Fig. \ref{fig:vis}-3b. 
Similar to the principal components, the magnitude of each weight indicates the importance of the metric, while the sign suggests the direction of variation in the data distribution. 

As we apply PCA across the time dimension, we preserve the overall temporal variation at the expense of losing time dependencies. Every DR method also inevitably introduces inherent distortions in the embedding~\cite{jeon2021measuring}.
While such information loss may be viewed as a limitation of our visualization system, we consider this a necessary compromise to derive node clusters from large-scale raw performance data at interactive speed. 
Once the system user locates a node cluster of interest for further analysis, our intra-cluster analysis takes a different approach to examine the raw performance data without losing detailed spatial and temporal patterns.

\subsubsection{Intra-cluster analysis}
Along with comparing groups of nodes, our visualization system allows users to compare nodes that constitute a cluster using multi-resolution dynamic mode decomposition (mrDMD) coupled with frequency isolation using the mrDMD spectrum~\cite{BRUNTON20161}. mrDMD extracts dynamic structures from time-series datasets by reducing high-dimensional data into coupled spatial and temporal features \cite{mrdmd}. We first use the time series of the desired cluster from the Node Similarity view to extract the mrDMD spatiotemporal modes. We then extract the high-power mrDMD modes at a desired frequency range from the mrDMD spectrum~\cite{BRUNTON20161, multifidelity, incmrdmd} (Fig. \ref{fig:workflow}-II). This step removes noise from our results and retains frequencies of interest. We then compare the relative change (or z-scores) in these mrDMD modes from mrDMD modes computed similarly for baseline readings.
These baseline readings represent normal system behavior for a particular metric. We first extract these baseline readings by taking the interquartile range (IQR) of the measurements across all selected nodes and finding the longest continuous time window where all values fall within this range. We then duplicate this time window across the entire time series to simulate a baseline (i.e., normal) behavior. We compute the z-scores of these mrDMD modes from the baseline mrDMD modes. These z-scores indicate how far each node deviates from the baselines for a particular metric, which can reveal anomalous node and metric pairs when compared to the corresponding baseline behavior (\textbf{R3}). Users can then identify node behaviors with respect to their baseline within each cluster. The baselines are pre-computed or user-selected, allowing for customizable exploration of the evolving temporal behavior of these complex large-scale system dynamics.

Users can identify ``correlated'' metrics in the heatmap of z-scores when two metrics in the same node cluster show extremely high/low z-scores for a single node (\textbf{R2}). For instance, a negative correlation between two metrics is shown if a node has a high z-score for one metric and a low z-score for another metric. Similarly, if two metrics have high z-scores for a single node, that could indicate a positive correlation. 
These mrDMD z-score correlations are computed using metric-specific baselines. Because baseline behavior is not explicitly labeled in our datasets and our statistical method for extracting it can vary across metrics, we allow users to manually adjust baselines in the UI (see Fig. \ref{fig:interact}). This flexibility ensures consistency in the analysis across different node clusters, where normal behavior may differ due to workload and resource allocation variations.

\subsection{Visualization Module}

To support interactive visual analysis of HPC system monitoring data, we design four coordinated visualization components in our system, as shown in Fig. \ref{fig:vis}.
Below, we introduce these components following the analysis workflow.

The \textbf{Time Domain} view provides a temporal overview of compute node activity using a stacked bar chart (Fig. \ref{fig:vis}-1). 
Each bar represents a segment in which a node showed null readings across all metrics for a single time point.  
Users can select a time window for detailed inspection in the Metric Reading view.
This view assists users in identifying relevant time periods to investigate individual node and cluster behaviors and observe changes and trends leading up to system failure events (e.g., 21:00, Fig. \ref{fig:vis}-1).

The \textbf{Node Similarity} view summarizes relationships among compute nodes extracted from inter-cluster analysis. 
Each point represents the behavior of a single compute node across all metrics and time points, where proximity indicates similar behavior. As shown in Fig. \ref{fig:vis}-2, the compute nodes form four distinct clusters after applying the MultiDR framework--suggesting four distinct patterns in node behaviors for this particular day.
To visually and mathematically distinguish groups of nodes, %highlight distinct node clusters more effectively, 
we perform $k$-means clustering on the points and color the nodes based on their assigned centroid. Users can configure the number of centroids ($k$) in the Node Similarity view. 
Users can also isolate node cluster(s) of interest via a lasso selection, prompting a detailed cluster interpretation/comparison in the Metric Reading view.

\begin{figure*}[ht]
    \centering
    \includegraphics[width=0.9\textwidth]{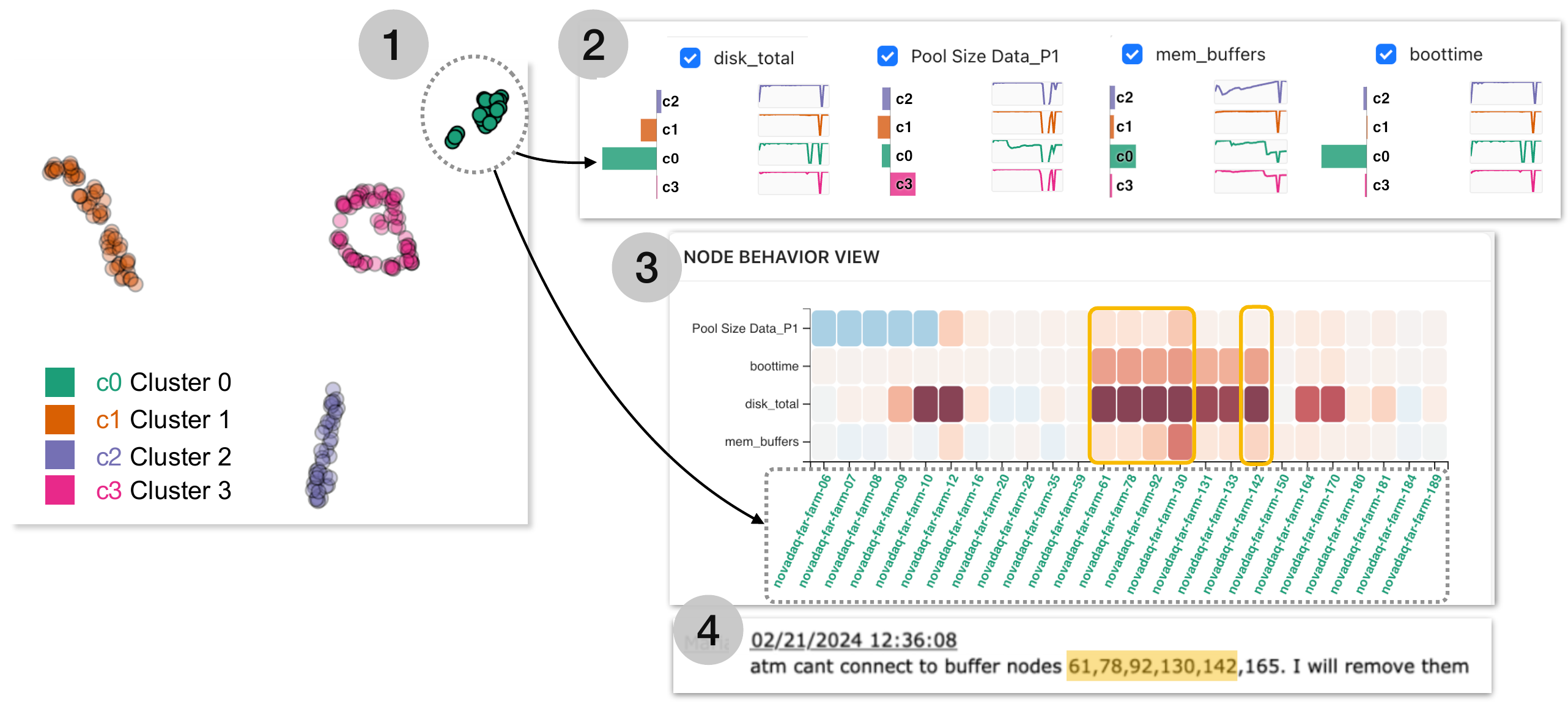}
    \caption{Intra-cluster analysis results of cluster 0 in the Ganglia monitoring dataset. Four highly-contributing metrics were selected for analysis (2). Differences in z-scores (3) are observed between a smaller sub-cluster of nodes (06-10) and the rest of \texttt{c0}. A subset of nodes 
    (61, 78, 92, 130, 142) was also noted in a log book entry during an \textit{unplanned downtime} event (4).}
    \label{fig:casestudy1}
\end{figure*} 

The \textbf{Metric Reading} view contains three components: metric selection panel (Fig. \ref{fig:vis}-3a), a cluster-based reading summary (Fig. \ref{fig:vis}-3b), and a reading inspection component (Fig. \ref{fig:vis}-3c).
The metric selection panel lists all available metrics, each accompanied by a horizontal bar chart that indicates its contribution to cluster uniqueness (Fig. \ref{fig:vis}-3a).
Each bar represents a cluster (e.g., \texttt{c0}-\texttt{c3} for $k$=4); its length reflects the metric's importance (ccPCA), while its direction suggests the variations in data values (left: higher; right: lower).
We rank the metrics by their absolute contribution values to efficiently identify metrics of interest. Users can also filter and search the list for specific metrics. 
As shown in Fig. \ref{fig:vis}-3, ``mem free'' contributes strongly to characterizing cluster \texttt{c0}, while ``CPU wio'' contributes strongly to cluster \texttt{c2}.

We also include the average polylines for each cluster (Fig. \ref{fig:vis}-3b), which represents the temporal average of the metric values across all nodes in the cluster. We apply a smoothing window for easier interpretation of short-term fluctuations. This allows users to quickly interpret cluster differences and low-level temporal patterns. For instance, we see the effects of the system failure from 21:00 (Fig. \ref{fig:vis}-1) in the spikes in ``CPU wio'' and dips in the ``load one'' metric (Fig. \ref{fig:vis}-3b). 
This view also guides users in selecting important metrics for raw reading inspection. In the reading inspection component (Fig. \ref{fig:vis}-3c), users can examine raw readings of the selected nodes and brush over the charts to specify the ``normal'' range. These user-selected readings, as one input to intra-cluster analysis, serve as a baseline to identify nodes with dissimilar readings as anomalies (\textbf{R4}).

The \textbf{Node Behavior} view presents anomalies detected from intra-cluster analysis in a heatmap (Fig. \ref{fig:vis}-4). 
Rows represent selected metrics from the Metric Reading view, columns correspond to selected nodes from the Node Similarity view, and cells show z-scores indicating the deviation from user-selected baselines. For instance, the nodes in cluster \texttt{c2} (Fig. \ref{fig:vis}-4) exhibit higher z-scores in the ``load five", ``CPU idle", and ``CPU system" metrics, suggesting abnormal behavior in comparison to the nodes in cluster \texttt{c3}. Users can hover over cells to highlight corresponding raw readings in the Metric Reading view for validation.

Together, these interactive visual components support seamless transitions between inter- and intra-cluster analysis (\textbf{R5}), enabling users to iteratively explore node clusters, interpret their characteristics, and detect potential anomalous nodes while abstracting out complex computations of the analysis module (Section \ref{sec:methodology}-B).

\section{Case Studies}
To evaluate the capabilities of our system in analyzing raw, large-scale monitoring data, we conduct %two 
case studies on two HPC monitoring datasets, obtained from extreme-scale distributed computing infrastructure. 
We show the results of each analysis step, and how interactions with each view show hardware anomalies and enable explanatory DR. 

\subsection{Ganglia Logs}
\textbf{Dataset.} This study is part of the Tachyon Project \cite{tachyon}, which explores new approaches to modeling the infrastructure involved in large-scale scientific computing workflows associated with the Deep Underground Neutrino Experiment (DUNE). Data generated by DUNE is vast, containing monitoring data from simulated HEP (High Energy Physics) experiments collected using a Ganglia-based Data Acquisition Monitoring System deployed at Fermilab (FNAL) \cite{nova}. Ganglia \cite{ganglia} provides near real-time monitoring and performance metrics for computer networks. The 206 compute nodes receive data every 5 milliseconds, and Ganglia buffers aggregated performance metrics every 15 seconds. 
System experts also maintain a logbook of system events. %in a shared PostgreSQL database. 
Each entry includes a timestamp, affected subsystem, description, and additional user comments. We focus on log events tagged as ``unplanned downtime" to validate the visual analytics results and ensure anomalies associated with problematic hardware components are captured.

\vspace{0.05in}

\noindent
\textbf{Analysis.} For preprocessing, we converted the Round Robin Database (RRD) \cite{rrd} files to comma-separated-value (CSV) files, combining 31 time-series performance metrics from all 206 nodes. We applied MulTiDR to one day of data (195 active nodes, 31 metrics, 4,416 timestamps). The first DR step used PCA across the time domain for each node, and the second DR step applied UMAP (\textit{n\_components}=2, \textit{n\_neighbors}=15, \textit{min\_dist}=0.1, \textit{random\_state}=42) across the metric domain. The resulting 2D projections were clustered using $k$-means ($k$=4). 
Bar plots (Fig. \ref{fig:vis}-3a) show each metric's contribution per cluster. 
For example, ``load one" strongly contributes to the uniqueness of cluster 1 (orange), and we see it's polyline experience a dramatic dip around 18:00 when nodes from clusters 0, 1, and 3 show downtime (Fig. \ref{fig:vis}-1). 
Users can identify periods of interest via the Time Domain view (Fig. \ref{fig:vis}-1) and select nodes from affected clusters. The temporal view (Fig. \ref{fig:vis}-2) highlights two distinct events: 18:00 (affecting clusters 0, 1, and 3) and 21:00 (all clusters showing nodes with null readings, indicating \textit{planned} downtime). The first event aligns with a log entry for unplanned downtime at 18:26:33, lasting approximately 25 minutes. Looking at cluster 0, UMAP projections reveal a smaller subgroup of nodes separated from the main cluster (Fig. \ref{fig:casestudy1}-1). 
We select metrics contributing highly to the DR results (``disk total", ``mem buffers", ``boot time") or show unique polyline behavior from other clusters (``Pool Size Data\_P1") (Fig. \ref{fig:casestudy1}-2). 
Many of these nodes exhibit varying z-scores, distinguishing nodes within the same assigned cluster (Fig. \ref{fig:casestudy1}-3). 

Our visualization system also identified a group of nodes that became unresponsive and were subsequently removed from a job run. A log book entry at 12:36:08 notes that buffer nodes 61, 78, 92, 130, 142, and 165 were unresponsive and excluded from execution (Fig. \ref{fig:casestudy1}-4). With the exception of node 165, all of these nodes were assigned to cluster 0 and showed elevated z-scores for the ``disk total" and ``boot time" metrics. In contrast, node 165 was grouped into cluster 1, and showed stable readings across all metrics, consistent with the rest of the cluster. 
Variation within this group can be further explored in the Metric Reading view by inspecting raw time series for individual nodes. For instance, node 130 displays higher z-scores across several metrics compared to its peers, while nodes 131 and 133--not flagged in the log book--exhibit similar anomalous behavior, warranting additional investigation.

\begin{figure}[h]
    \centering
    \includegraphics[width=\linewidth]{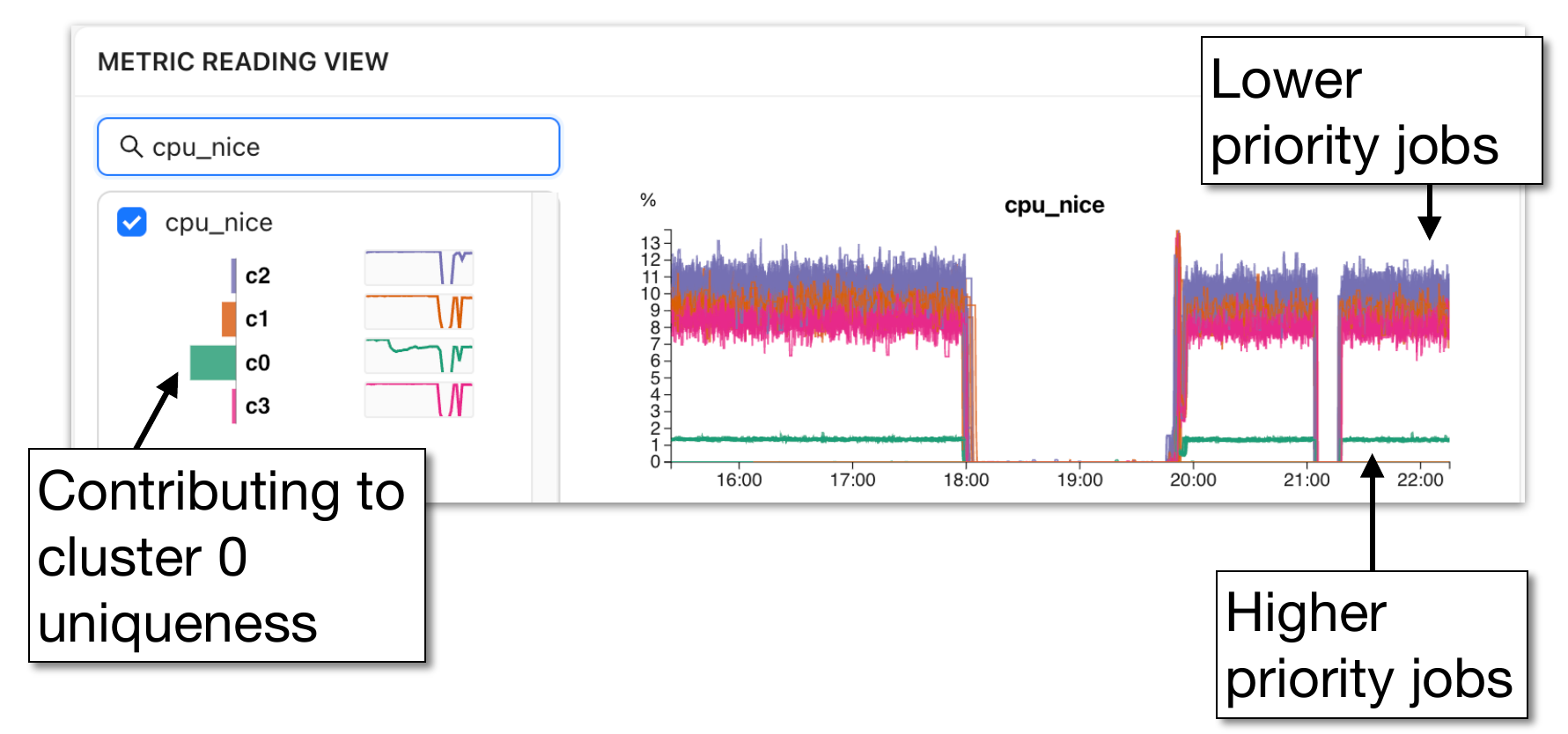}
    \vspace{-0.15in}
    \caption{Inter-cluster analysis results of Ganglia logs for ``CPU nice." Lower readings indicate the CPU time occupied by jobs with higher scheduling priority. This metric effectively captures the inter-cluster differences between \texttt{c0}-\texttt{c3}.}
    \label{fig:casestudy1-3}
    \vspace{-0.1in}
\end{figure}

Additionally, our visualization identified inter-cluster differences based on job priority (Fig. \ref{fig:casestudy1-3}). Our DR clustering separated nodes from cluster 0, which reflect much lower ``CPU nice" readings compared to the other clusters. We see from the Metric Reading view that ``CPU nice" contributed highly to cluster 0, and the raw time series reveals the nodes in cluster 0 were assigned to much higher priority jobs. 

Overall, our intra-cluster analysis of cluster 0 enabled us to (i) isolate an inactive subgroup of nodes (see Fig. \ref{fig:casestudy1}-1), (ii) detect an anomalous node within an otherwise stable cluster (see Fig. \ref{fig:casestudy1}-4), and (iii) identify a set of problematic nodes linked to a single job run (see Fig. \ref{fig:casestudy1}-3). These findings demonstrate how the system facilitates multi-level anomaly detection, bridging high-level cluster patterns with node-level diagnostics to support both immediate troubleshooting and deeper node analysis. 

\subsection{Environment Logs}
\textbf{Dataset.} The second dataset consists of environment log data collected from Theta, an 11.7 Petaflops Cray XC40 supercomputer system, previously analyzed in related studies \cite{mela}. 
These environment logs, also referred to as SEDC (System Environment Data Collections) logs \cite{9825952}, are recorded at 10–30 second intervals and grow rapidly, reaching terabytes of data within weeks. These logs capture sensor readings across multiple system levels—including rack, blade, and node—covering measurements such as power consumption, air and water temperature, and voltages. 
The environment logs used in this study comprise 4,392 Intel Xeon Phi-based compute nodes organized into 24 racks, with approximately 150 sensor readings per node collected at each interval. The logs were queried from an InfluxDB time series database and exported to CSV format for preprocessing and visualization. To validate the results of our analysis, we also incorporate job logs from the same system and timeframe, which provide metadata including timestamps, users, job categories, nodes utilized, and job exit statuses.

\begin{figure}[h]
    \centering
    \includegraphics[width=\columnwidth]{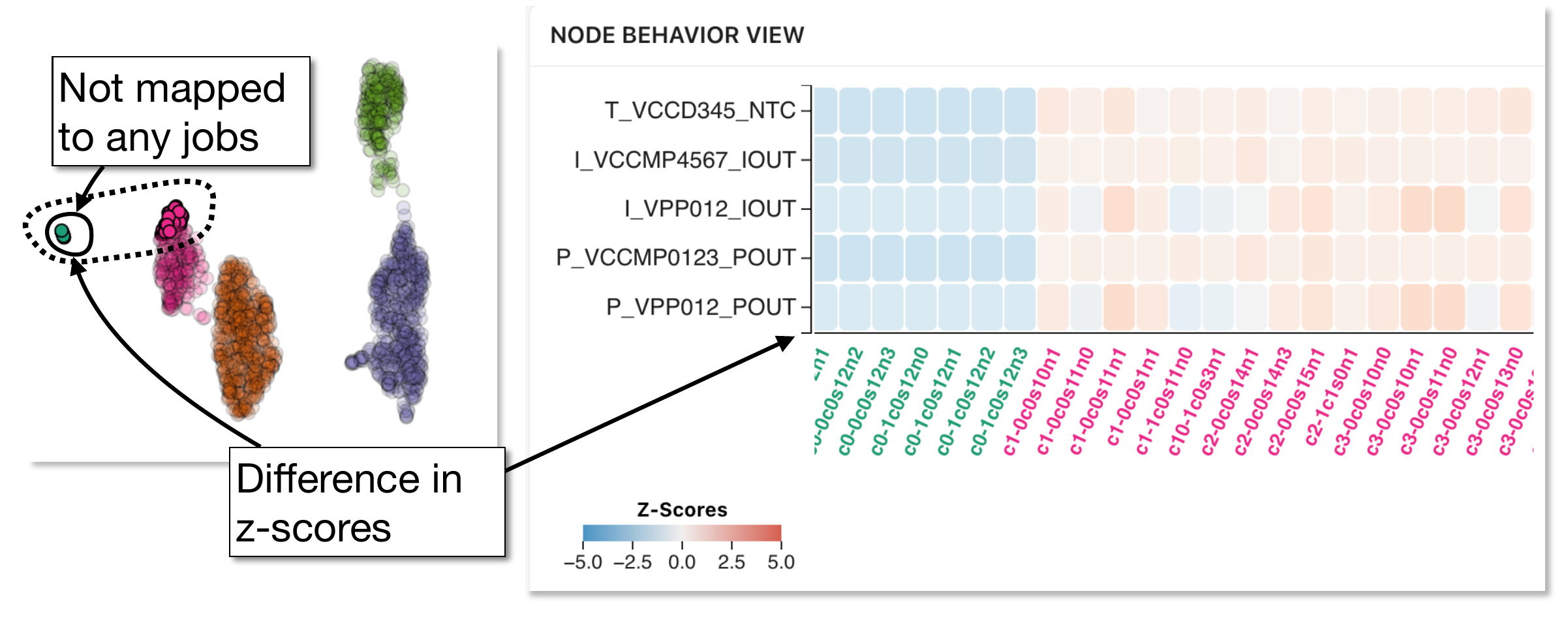}
    \vspace{-0.2in}
    \caption{Inter-cluster analysis results for clusters 1 and 2 (\texttt{c1}, \texttt{c2}) of the environment logs. Nodes in \texttt{c2} are not mapped to any jobs. Shown in the z-scores heatmap, nodes from \texttt{c2} have noticeably lower z-scores across all selected features.}
    \label{fig:casestudy2-1}
    \vspace{-0.1in}
\end{figure}
\begin{figure}[h]
    \centering
    \includegraphics[width=\columnwidth]{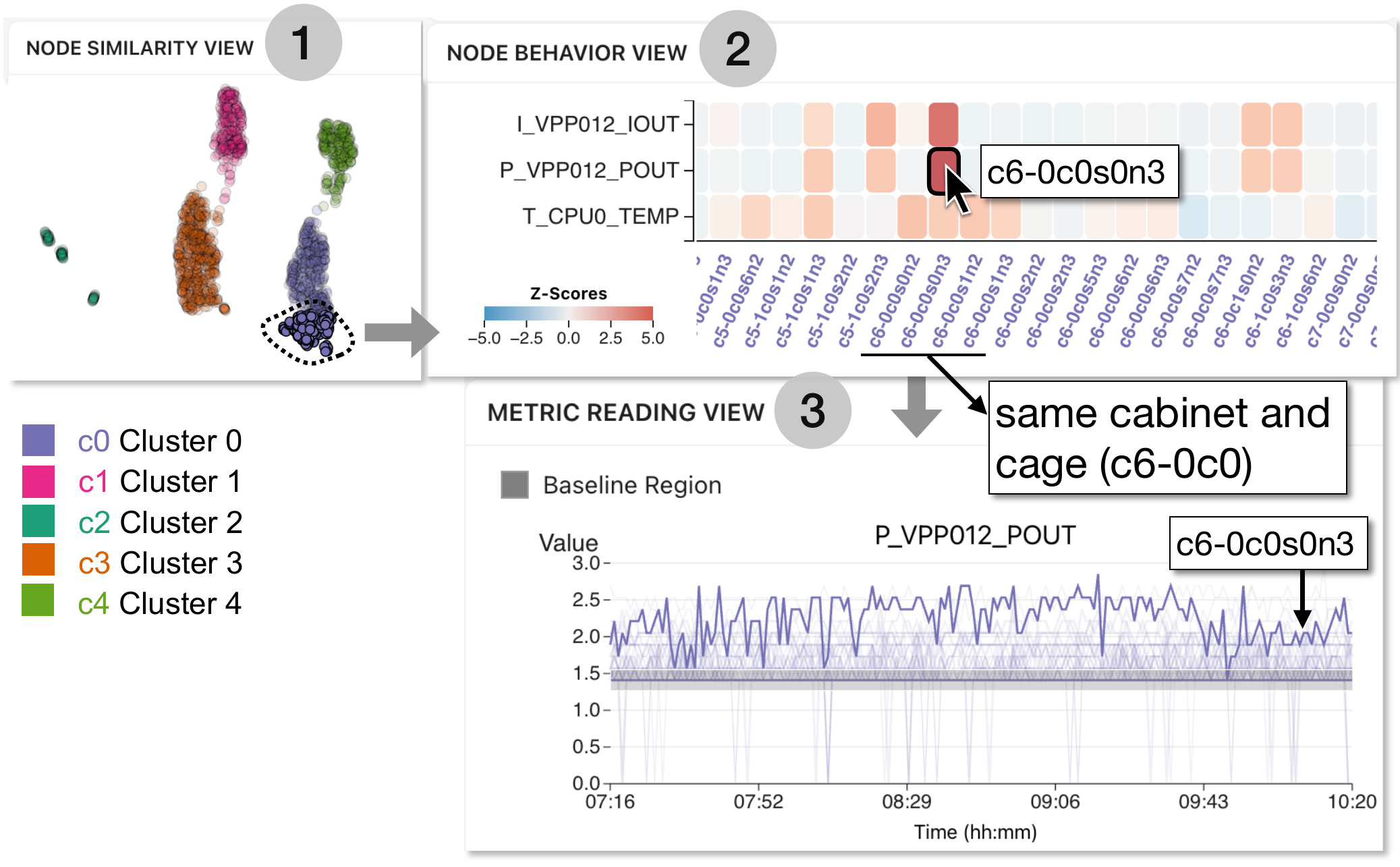}
    \vspace{-0.1in}
    \caption{Intra-cluster analysis results for cluster 0. We compare the z-scores (2) between nodes in \texttt{c0} (1). We notice correlations between power and current. Four nodes from the same cabinet and cage (\texttt{c6-0c0}) have high CPU0 temperature readings. Hovering over node \texttt{c6-0c0s0n3}, we see the values sitting above the baseline for power and current (3).}
    \label{fig:casestudy2-2}
    \vspace{-0.1in}
\end{figure}

\vspace{0.05in}

\noindent
\textbf{Analysis.} We preprocessed the environment logs by combining the node-level time-series for half a day of data (2018-06-09 00:00:00 - 2018-06-09 13:19:00). This subset included 1,600 nodes, 123 metrics, and 1,096 timestamps per metric. We applied the MulTiDR framework \cite{multidr}: PCA across the time domain for each node, followed by UMAP (\textit{n\_components}=2, \textit{n\_neighbors}=50, \textit{min\_dist}=0.5, \textit{random\_state}=42) across the metric domain. The final 2D projections revealed four large clusters and two smaller ones. Using $k$-means, we set $k=5$, and subsequently applied ccPCA \cite{ccpca} to compare the distinguishing metric characteristics of the clusters. The DR results for the preprocessed environment logs are depicted in Fig. \ref{fig:casestudy2-2}-1. 

Cluster 2, in particular, exhibits a consistently high rate of null values throughout the entire period, indicating that the cluster was formed due to persistent zeros across all metrics. Comparing clusters 1 and 2 (Fig. \ref{fig:casestudy2-1}), nodes in cluster 2 exhibit z-scores far below baseline for every selected feature.

Looking closer at intra-cluster patterns, we observed correlations between power and current metrics (\textit{P\_VPP012\_POUT} and \textit{I\_VPP012\_IOUT}), with several nodes displaying consistently higher z-scores for both (Fig. \ref{fig:casestudy2-2}-2). Within cluster 0, we identified four nodes from the same cabinet and cage (\texttt{c6-0c0}) with elevated CPU0 temperature readings. Examining the time series of node \texttt{c6-0c0s0n3}, we confirmed that both power and current values remained above baseline (Fig. \ref{fig:casestudy2-2}-3). 

This analysis highlights how our system can (i) isolate inactive node groups (Fig. \ref{fig:casestudy2-1}), (ii) reveal inter-cluster differences tied to job activity (Fig. \ref{fig:casestudy2-1}), and (iii) uncover intra-cluster anomalies and feature correlations (Fig. \ref{fig:casestudy2-2}). These results demonstrate the system’s ability to connect job-level activity with sensor-level behaviors, enabling both operational monitoring and exploratory diagnostics.
\section{Discussion}
We have demonstrated the analysis capabilities of our system for investigating HPC behavior in raw, large-scale monitoring data. 
Our approach uncovered node clusters with distinct characteristics, validated against environment logs and job logs, and enabled reasoning about anomalous nodes by integrating cluster-level patterns with deviations in node behavior.
In this section, we discuss  the strengths and limitations of our system by reviewing feedback from HPC experts and outlining our future work to enhance usability, scalability, and integration with production environments.

\begin{figure}[t]
    \centering
    \begin{subfigure}{\linewidth}
        \centering
        \includegraphics[width=\linewidth]{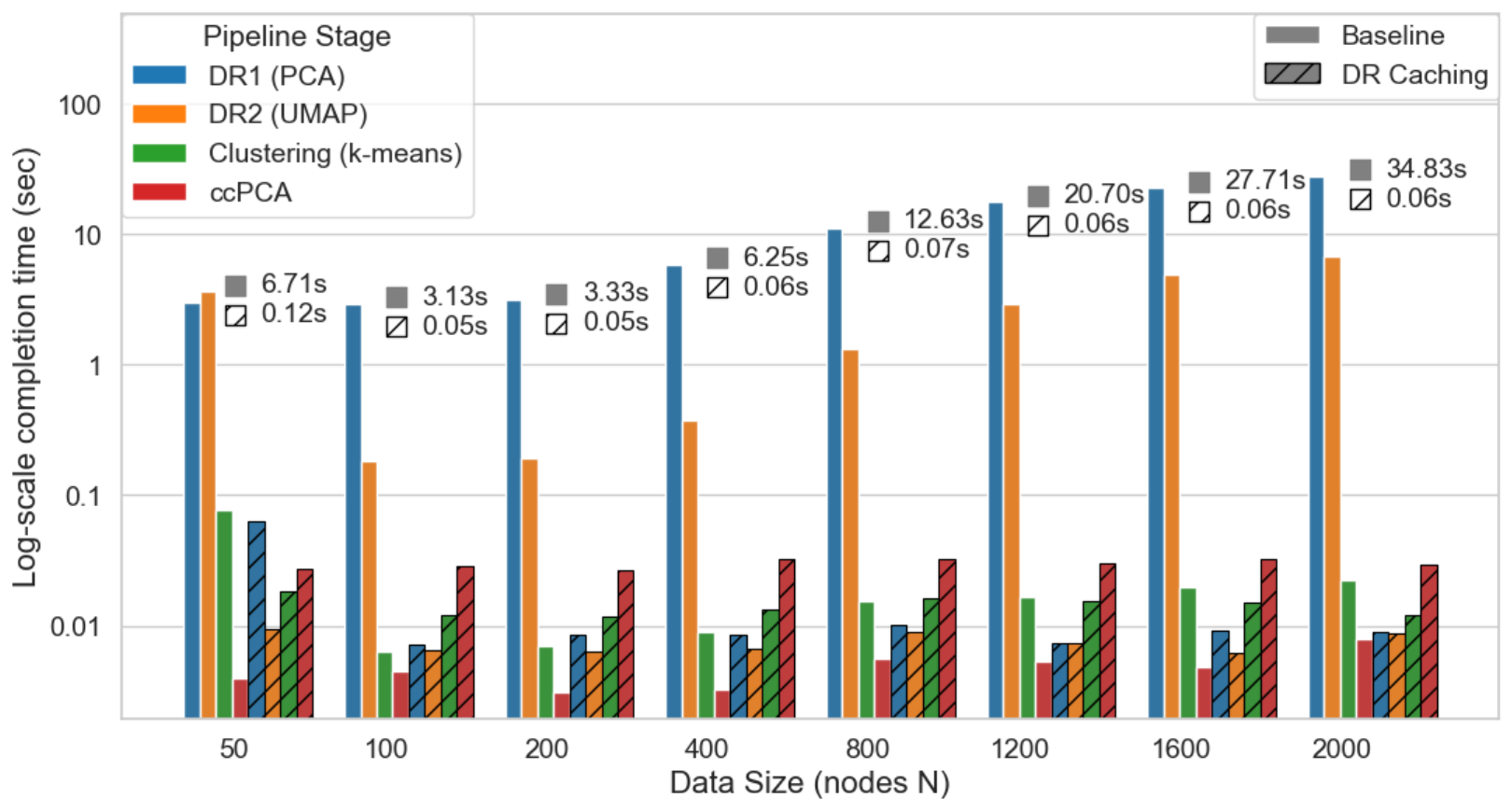}
        \caption{Scaling the number of nodes $N$, with fixed $M=46$.}
        \label{fig:scalability-nodes}
    \end{subfigure}

    \vspace{0.1in}

    \begin{subfigure}{\linewidth}
        \centering
        \includegraphics[width=\linewidth]{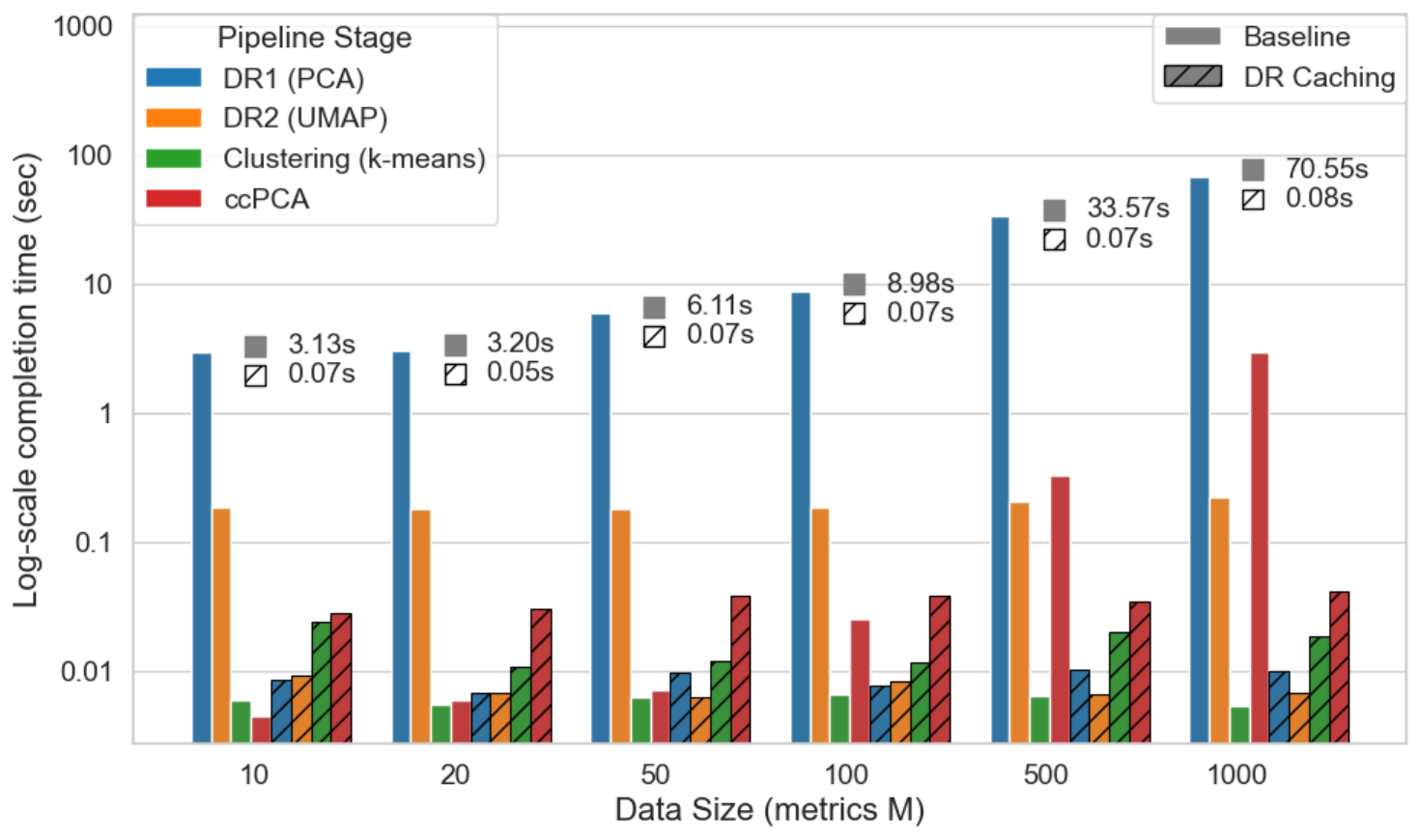}
        \caption{Scaling the number of metrics $M$, with fixed $N=195$.}
        \label{fig:scalability-metrics}
    \end{subfigure}

    \caption{Comparison of pipeline completion time, showing how performance scales with the number of nodes $N$ and metrics $M$ in the dataset and the improvement in caching the results of each DR step.}
    \label{fig:scalability}
\end{figure}

\subsection{Performance Evaluation}
We evaluate the performance of our clustering pipeline by testing data scalability (for both nodes and metrics), conducting a qualitative comparison of clustering across different DR methods, and testing cluster quality of the dimensionality reduction. All experiments were conducted on a 12-core Apple M2 Max CPU with 64\,GB of unified memory. For cluster quality and comparison, we use raw HPC monitoring data from Ganglia logs ($N=195$, $M=46$, $T=4,416$) and environment logs from the Cray XC40 supercomputer ($N=1,600$, $M=123$, $T=1,096$). 

\textbf{Scalability Study}. We evaluate the scalability of our pipeline along two dimensions: (i) the number of nodes $N$ with a fixed number of metrics ($M=46$), and (ii) the number of metrics $M$ with a fixed number of nodes ($N=195$). Since the explainable clustering algorithms ($k$-means, ccPCA) rely on the DR results, and the DR steps dominate runtime (both account for over 80--90\% of total computation time at large $N$), the overall baseline execution of our pipeline does not scale well when increasing the number of nodes and metrics in the dataset (shown in Fig. \ref{fig:scalability}). To address these limitations, our system caches the results of each DR step, allowing users to interactively re-run the clustering and mrDMD phases without recomputing embeddings. This reduces the runtime of the total compute pipeline by up to two to three orders of magnitude, decreasing from over one minute in the metric-scaling experiment (Fig. \ref{fig:scalability-metrics}) and over 30\, seconds in the node-scaling experiment (Fig. \ref{fig:scalability-nodes}) to under 10\,ms. 
As a result, the cost of DR1 and DR2 becomes effectively independent of $N$ and $M$. Additionally, since the clustering operates on low-dimensional embeddings, and ccPCA and mrDMD operate on a fixed set of components, their computational overhead remains minimal once DR results are reused.

\begin{table*}[t]
\centering
\caption{Cluster quality metrics for different DR pipelines.
Higher values are better for $\uparrow$ metrics and lower values are better for $\downarrow$ metrics. \textbf{Bolded} values represent our pipeline's default method.}
\label{tab:cluster-quality}
{\small
\begin{tabular}{lccccc}
\toprule
\textbf{DR Method} &
\textbf{Silhouette} $\uparrow$ &
\textbf{Davies--Bouldin} $\downarrow$ &
\textbf{Calinski--Harabasz} $\uparrow$ &
\textbf{Trustworthiness} $\uparrow$ &
\textbf{Continuity} $\uparrow$ \\
\midrule

\multicolumn{6}{l}{\textit{Ganglia logs ($N=195$, $M=46$, $T=4,416$)}} \\
\midrule
PCA   & 0.418 & 0.881 & 0.0 & 1.0 & 0.86 \\
\textbf{UMAP}  & \textbf{0.755} & \textbf{0.289} & \textbf{0.207} & \textbf{1.0} & \textbf{0.609} \\
\textsc{t-SNE}  & 0.514 & 0.684 & 0.121 & 1.0 & 0.7 \\
LDA  & 0.751 & 0.484 & 1.0 & 1.0 & 0.542 \\
ULCA  & 0.925 & 0.165 & 0.524 & 1.0 & 0.82 \\
TULCA  & 0.818 & 0.397 & 0.0 & 1.0 & 0.71 \\
\midrule
\multicolumn{6}{l}{\textit{Environment logs ($N=1,600$, $M=123$, $T=1,096$)}} \\
\midrule
PCA   & 0.940 & 0.104 & 1.0 & 1.0 & 0.997 \\
\textbf{UMAP}  & \textbf{0.692} & \textbf{0.381} & \textbf{0.612} & \textbf{0.998} & \textbf{0.862} \\
\textsc{t-SNE}  & 0.591 & 0.545 & 0.202 & 1.0 & 0.747 \\
LDA  & 0.583 & 0.717 & 1.0 & 1.0 & 0.414 \\
ULCA  & 0.35 & 0.709 & 0.504 & 1.0 & 0.859 \\
TULCA  & 0.508 & 0.550 & 0.0 & 1.0 & 0.507 \\
\bottomrule
\label{tab:clustering-results}
\end{tabular}
}
\vspace{-0.2in}
\end{table*}

\begin{figure}[ht]
    \centering
    \includegraphics[width=\columnwidth]{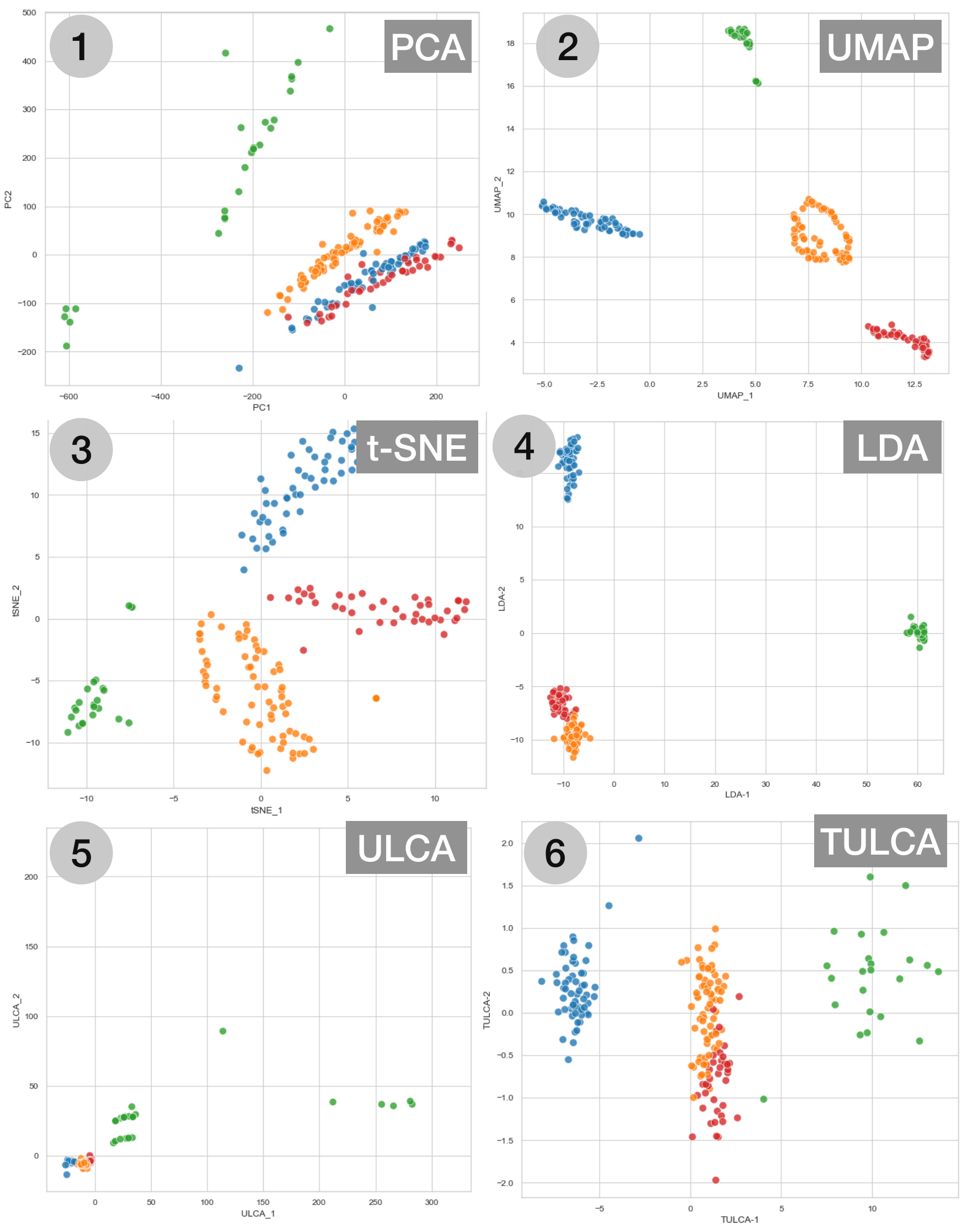}
    \vspace{-0.1in}
    \caption{Comparison of DR methods for node-level system behavior. We compare unsupervised linear (1), unsupervised nonlinear (2), (3), supervised linear (4), contrastive linear (5), and tensor-based discriminative (6) methods. 
    }
    \label{fig:cluster-comp}
    \vspace{-0.1in}
\end{figure}

\textbf{Qualitative Comparison.} To evaluate how different dimensionality reduction strategies behave on large-scale system monitoring data, we qualitatively compare our embeddings against a representative set of baseline methods. 
\textsc{PCA} \cite{pca}, \textsc{UMAP} \cite{umap}, and \textsc{t-SNE} \cite{tsne} are commonly used unsupervised baselines in system analysis and visualization. \textsc{LDA} \cite{lda} provides a supervised linear baseline that explicitly maximizes class separability, and \textsc{ULCA} \cite{ulca} extends this formulation by incorporating contrastive objectives. Finally, \textsc{TULCA} \cite{tulca} represents a tensor-based generalization that jointly models temporal and metric dimensions. Fig. \ref{fig:cluster-comp} compares the resulting two-dimensional embeddings produced by each method. Nodes are colored according to cluster labels obtained by applying $k$-means ($k=4$) on our default embedding (PCA$\rightarrow$\textsc{UMAP}), shown in Fig. \ref{fig:cluster-comp}-2. These labels are held fixed across all methods to facilitate direct visual comparison. The axes represent the first two components of each dimensionality reduction method. Across methods, we observe clear differences in cluster separation and structural preservation. \textsc{UMAP} produces the most distinct and well-separated clusters, while \textsc{PCA}, \textsc{LDA}, and \textsc{TULCA} exhibit noticeable overlap between the red and orange clusters, indicating reduced separability under linear or globally optimized projections. 

\textbf{Cluster Quality}. We evaluate the quality of clusters produced by different DR methods using a combination of clustering metrics—silhouette score~\cite{rousseeuw_silhouettes_1987}, Davies--Bouldin index~\cite{davies_cluster_1979}, and Calinski--Harabasz index~\cite{calinski_dendrite_1974}—together with label trustworthiness and continuity to assess neighborhood preservation in the resulting embeddings~\cite{jeon_classes_2024}. The silhouette score (range $[0,1]$) captures cluster separation and cohesion, while the Davies--Bouldin index (range $[0,1]$) measures relative intra- versus inter-cluster dispersion, and the Calinski--Harabasz index (range $[0,\infty]$, normalized in Table \ref{tab:clustering-results}) quantifies variance separation between clusters. The label trustworthiness (range $[0,1]$) suggests how well the local neighborhood is preserved after applying the DR methods, while label continuity (range $[0,1]$) suggests how well the global structure is retained in the low-dimensional space. 
Results for evaluating the cluster quality across DR methods for the Ganglia and environment log datasets are summarized in Table \ref{tab:clustering-results}. 
Although several DR methods achieve strong cluster quality scores, the \textsc{UMAP} configuration used in our system exhibits consistently robust performance across datasets and evaluation metrics. \textsc{UMAP} substantially outperforms purely linear pipelines across all cluster quality measures, while avoiding the sensitivity and stochastic variability often observed in \textsc{t-SNE}-based pipelines. 
Note that we do not compare the mrDMD result quality as it is covered in the original paper~\cite{incmrdmd}.

\subsection{Expert Feedback}
To evaluate the usefulness of our system, we conducted one-hour interviews with two experts (\textbf{E1} and \textbf{E2}), who represent our system target users. 
We presented Case A to \textbf{E1} and Case B to \textbf{E2}, as they have direct experience with the corresponding systems ($10$ and $20$ years, respectively).
Both experts confirmed that our system provides valuable support for monitoring and reasoning about HPC behaviors. 
Regarding visualization components, \textbf{E1} noted that the Time Domain view effectively highlights the nodes responsible for crashes, while \textbf{E2} praised the Node Behavior view and the Metric Reading view for their intuitive design. 
For analysis capabilities, \textbf{E1} emphasized that the system could assist in detecting misconfigured compute nodes--a frequent problem where fine-tuned node configurations are unintentionally overwritten during setup. \textbf{E2} appreciated the ability to define the baseline behaviors for compute nodes, noting that this feature provides better support for monitoring.
Both experts also suggested directions for improvement. \textbf{E2} recommended incorporating metadata about monitoring metrics to support more informed metric selection, while \textbf{E1} expressed interest in analyzing the early formation of node clusters prior to system crashes.
Finally, both experts saw strong potential for applying our system to real-time monitoring scenarios. Their feedback will guide the next design iteration, with a focus on enhancing interpretability, proactive anomaly detection, and real-time applicability.

\subsection{Future Work}
Our current system is not without limitations, suggesting directions for future work, particularly addressing the data preprocessing and real-time analysis tasks. While our design  can effectively integrate both offline and online analysis methods, the present work focuses on retrospective monitoring. We are working %plan
to extend our pipeline toward fully end-to-end, real-time processing by directly querying monitoring databases and updating the visualizations incrementally as new data arrives. Both expert users highlighted streaming support as an important next step, motivating us to explore incremental algorithms that are optimized to handle continuous data flows.
We also plan to deploy our system on cloud platforms to support broader accessibility and real-world evaluation in operational settings. This will enable us to collect richer expert feedback, assess scalability across diverse HPC workloads, and refine the usability of the system in live monitoring scenarios. 
While the case studies in this work focus on core compute metrics, the framework’s underlying tensor-based architecture ($Node \times Metric \times Time$) is generic. Future iterations will incorporate heterogeneous telemetry, such as network throughput and I/O storage logs, to provide a more holistic view of system health across the entire hardware stack.
\section{Conclusion}
The growing complexity and volume of large-scale 
HPC system monitoring data demands new inspecting and analytical reasoning capabilities. 
We present a visual analytics tool that abstracts complex cluster-based algorithms into interactive visualizations for exploring HPC monitoring data at both inter-cluster and intra-cluster levels. 
The tool provides scalable overviews of large-scale system behavior while supporting detailed inspection, enabling analysts to refine results using their domain expertise.
We evaluated our approach using two production HPC monitoring datasets and validated the findings using both job log data, log book entries, and expert feedback. With the tool, we were able to effectively identify  problematic nodes, inactive node groups, metric correlations, cluster differences, and within-cluster anomalies. Expert interviews confirmed its usability and highlighted its potential for monitoring and diagnosis in real-world settings. 
Our work contributes to the development of robust, interpretable, and scalable solutions for HPC monitoring analysis.

\section*{Acknowledgments}
This research is supported in part by the U.S. Department of Energy with grant DE-SC0024580. This research used data and computational resources of the Argonne Leadership Computing Facility (ALCF), including the Theta system, a U.S. Department of Energy Office of Science User Facility supported under Contract No. DE-AC02-06CH11357. S., V.V., and M.E.P. contributions were supported in part under the same contract.

\bibliographystyle{IEEEtran}
\bibliography{00_main}

% \vspace{12pt}
% \color{red}
% IEEE conference templates contain guidance text for composing and formatting conference papers. Please ensure that all template text is removed from your conference paper prior to submission to the conference. Failure to remove the template text from your paper may result in your paper not being published.

\end{document}